\documentclass{aastex}
\usepackage{amsmath}
\usepackage{amssymb}

\shorttitle{Back-tracing and flux reconstruction for solar events with PAMELA}
\shortauthors{Bruno et al.}

\begin{document}

\title{Back-Tracing and Flux Reconstruction for Solar Events with PAMELA}

\author{
A.~Bruno$^{1,*}$,
O.~Adriani$^{2,3}$,
G.~C.~Barbarino$^{4,5}$,
G.~A.~Bazilevskaya$^{6}$,
R.~Bellotti$^{1,7}$,
M.~Boezio$^{8}$,
E.~A.~Bogomolov$^{9}$,
M.~Bongi$^{2,3}$,
V.~Bonvicini$^{8}$,
S.~Bottai$^{3}$,
U.~Bravar$^{10}$,
F.~Cafagna$^{7}$,
D.~Campana$^{5}$,
R.~Carbone$^{8}$,
P.~Carlson$^{11}$,
M.~Casolino$^{12,13}$,
G.~Castellini$^{14}$,
E.~C.~Christian$^{15}$,
C.~De~Donato$^{12,17}$,
G.~A.~de~Nolfo$^{15}$,
C.~De~Santis$^{12,17}$,
N.~De~Simone$^{12}$,
V.~Di~Felice$^{12,18}$,
V.~Formato$^{8,19}$,
A.~M.~Galper$^{16}$,
A.~V.~Karelin$^{16}$,
S.~V.~Koldashov$^{16}$,
S.~Koldobskiy$^{16}$,
S.~Y.~Krutkov$^{9}$,
A.~N.~Kvashnin$^{6}$,
M.~Lee$^{10}$,
A.~Leonov$^{16}$,
V.~Malakhov$^{16}$,
L.~Marcelli$^{12,17}$,
M.~Martucci$^{17,20}$,
A.~G.~Mayorov$^{16}$,
W.~Menn$^{21}$,
M.~Merg\`e$^{12,17}$,
V.~V.~Mikhailov$^{16}$,
E.~Mocchiutti$^{8}$,
A.~Monaco$^{1,7}$,
N.~Mori$^{2,3}$,
R.~Munini$^{8,19}$,
G.~Osteria$^{5}$,
F.~Palma$^{12,17}$,
B.~Panico$^{5}$,
P.~Papini$^{3}$,
M.~Pearce$^{11}$,
P.~Picozza$^{12,17}$,
M.~Ricci$^{20}$,
S.~B.~Ricciarini$^{3,14}$,
J.~M.~Ryan$^{10}$,
R.~Sarkar$^{22,23}$,
V.~Scotti$^{4,5}$,
M.~Simon$^{21}$,
R.~Sparvoli$^{12,17}$,
P.~Spillantini$^{2,3}$,
S.~Stochaj$^{24}$,
Y.~I.~Stozhkov$^{6}$,
A.~Vacchi$^{8}$,
E.~Vannuccini$^{3}$,
G.~I.~Vasilyev$^{9}$,
S.~A.~Voronov$^{16}$,
Y.~T.~Yurkin$^{16}$,
G.~Zampa$^{8}$,
N.~Zampa$^{8}$,
and V.~G.~Zverev$^{16}$.
}

\affil{$^{1}$ Department of Physics, University of Bari, I-70126 Bari, Italy.}
\affil{$^{2}$ Department of Physics and Astronomy, University of Florence, I-50019 Sesto Fiorentino, Florence, Italy.}
\affil{$^{3}$ INFN, Sezione di Florence, I-50019 Sesto Fiorentino, Florence, Italy.}
\affil{$^{4}$ Department of Physics, University of Naples ``Federico II'', I-80126 Naples, Italy.}
\affil{$^{5}$ INFN, Sezione di Naples, I-80126 Naples, Italy.}
\affil{$^{6}$ Lebedev Physical Institute, RU-119991 Moscow, Russia.}
\affil{$^{7}$ INFN, Sezione di Bari, I-70126 Bari, Italy.}
\affil{$^{8}$ INFN, Sezione di Trieste, I-34149 Trieste, Italy.}
\affil{$^{9}$ Ioffe Physical Technical Institute, RU-194021 St. Petersburg, Russia.}
\affil{$^{10}$ Space Science Center, University of New Hampshire, Durham, NH, USA.}
\affil{$^{11}$ KTH, Department of Physics, and the Oskar Klein Centre for Cosmoparticle Physics, AlbaNova University Centre, SE-10691 Stockholm, Sweden.}
\affil{$^{12}$ INFN, Sezione di Rome ``Tor Vergata'', I-00133 Rome, Italy.}
\affil{$^{13}$ RIKEN, Advanced Science Institute, Wako-shi, Saitama, Japan.}
\affil{$^{14}$ IFAC, I-50019 Sesto Fiorentino, Florence, Italy.}
\affil{$^{15}$ Heliophysics Division, NASA Goddard Space Flight Ctr, Greenbelt, MD, USA.}
\affil{$^{16}$ National Research Nuclear University MEPhI, RU-115409 Moscow, Russia.}
\affil{$^{17}$ Department of Physics, University of Rome ``Tor Vergata'', I-00133 Rome, Italy.}
\affil{$^{18}$ Agenzia Spaziale Italiana (ASI) Science Data Center, I-00133 Rome, Italy.}
\affil{$^{19}$ Department of Physics, University of Trieste, I-34147 Trieste, Italy.}
\affil{$^{20}$ INFN, Laboratori Nazionali di Frascati, I-00044 Frascati, Italy.}
\affil{$^{21}$ Department of Physics, Universit\"{a}t Siegen, D-57068 Siegen, Germany.}
\affil{$^{22}$ Indian Centre for Space Physics, 43 Chalantika, Garia Station Road, Kolkata 700084, West Bengal, India.}
\affil{$^{23}$ Previously at INFN, Sezione di Trieste, I-34149 Trieste, Italy. }
\affil{$^{24}$ Electrical and Computer Engineering, New Mexico State University, Las Cruces, NM, USA.}

\altaffiltext{*}{Corresponding author. E-mail address: alessandro.bruno@ba.infn.it.}

\begin{abstract}
The PAMELA satellite-borne experiment is providing first direct measurements of Solar Energetic Particles (SEPs) with energies from $\sim$80 MeV to several GeV in near-Earth space.
Its unique observational capabilities include the possibility of measuring the flux angular distribution and thus investigating possible anisotropies related to SEP events.
This paper focuses on the analysis methods developed to estimate SEP energy spectra as a function of the particle pitch angle with respect to the Interplanetary Magnetic Field (IMF).
The crucial ingredient is provided by an accurate simulation of the asymptotic exposition of the PAMELA apparatus, based on a realistic reconstruction of particle trajectories in the Earth's magnetosphere.
As case study, the results of the calculation for the May 17, 2012 event are reported.
\end{abstract}

\section{Introduction}\label{Introduction}
Solar Energetic Particles (SEPs) are high energy particles associated with explosive phenomena occurring in the solar atmosphere,
such as solar flares and Coronal Mass Ejections (CMEs).
They can significantly perturb the Earth's magnetosphere producing a sudden increase in particle fluxes and, consequently, in the radiation exposure experienced by spacecrafts and their possible crew.
SEPs constitute a sample of solar material and provide important information about the sources of the particle populations,
and their angular distribution can be used to investigate details of the particle transport in the interplanetary medium.

SEP measurements are performed both by in-situ detectors on spacecrafts and by ground-based Neutron Monitors (NMs).
While the former are able to measure SEPs with energies below some tens of MeV,
the latter can only register the highest energy SEPs ($\gtrsim$ 1 GeV) during the so-called Ground Level Enhancements (GLEs).
A large energy gap exists between the two groups of observations.

New accurate measurements of SEPs are being provided by the PAMELA experiment \citep{SEP2006,MAY17PAPER}.
In particular, the instrument is able to detect SEPs in a wide energy interval ranging from $\sim$80 MeV up to several GeV, hence bridging the low energy data by space-based instruments and the GLE data by the worldwide network of NMs. In addition, PAMELA is sensitive to the particle composition and it is able to reconstruct the flux angular distribution, thus enabling a clearer and more complete view of the SEP events.

This paper reports the analysis methods developed for the estimate of SEP energy spectra with the PA\-ME\-LA apparatus, as a function of the particle asymptotic direction of arrival. As case study, the results used in the analysis of the May 17, 2012 solar event \citep{MAY17PAPER} are presented.

\section{The PAMELA Experiment}\label{The PAMELA experiment}
PAMELA is a space-based experiment designed for a precise measurement of the charged cosmic radiation in the kinetic energy range from some tens of MeV up to several hundreds of GeV \citep{Picozza,PHYSICSREPORTS}.
The Resurs-DK1 satellite, which hosts the apparatus, was launched into a semi-polar (70 deg inclination) and elliptical (350--610 km altitude) orbit on June the 15$^{th}$ 2006.
In 2010 it was changed to an approximately circular orbit at an altitude of about 580 km. The spacecraft is 3-axis stabilized; its orientation is calculated by an onboard processor with an accuracy better than 1 deg. Particle directions are measured with a high angular resolution ($\lesssim$ 2 deg).
Details about apparatus performance, proton selection, detector efficiencies and measurement uncertainties can be found elsewhere (see e.g. \citet{ProHe,SOLARMOD}).

\section{Geomagnetic Field Models}\label{Geomagnetic field models}
The analysis of SEP events described in this work relies on the IGRF-11 \citep{IGRF11} and the TS07D \citep{TS07D,TS07D2} models for the description of the internal and external geomagnetic field, respectively:
the former employs a global spherical harmonic implementation of the main magnetic field;
the latter is a high resolution dynamical model of the storm-time geomagnetic field in the inner magnetosphere, based on recent satellite measurements.
Consistent with the dataset coverage, the model is valid within the region delimited by the magnetopause (based on \citealt{SHUE}) and by a spherical surface with a radius of 30 Earth's radii (Re).
Solar Wind (SW) and IMF parameters are obtained from the high resolution (5-min) Omniweb database \citep{OMNIWEB}.
The TS07D model is more flexible and accurate with respect to all past empirical models in reconstructing the distribution of storm-scale currents, so it is particularly adequate for the study of SEP events.

\section{Back-tracing Techniques}
Cosmic Ray (CR) cutoff rigidities and asymptotic arrival directions (i.e. the directions of approach before encountering the Earth's magnetosphere) are commonly evaluated by simulations, accounting for the effect of the geomagnetic field on the particle transport (see e.g. \citealt{SMART_ETAL} and references therein).
Using spacecraft ephemeris data (position, orientation, time), and the particle rigidity ($R$ = momentum/charge) and direction provided by the tracking system, trajectories of all detected protons are reconstructed by means of a tracing program based on numerical integration methods \citep{TJPROG,SMART}, and implementing the afore-mentioned geomagnetic field models.

To reduce the computational time, geomagnetically trapped \citep{PAMTRAPPED} and most albedo \citep{ALBEDO} particles (originated from the interactions of CRs with the Earth's atmosphere) are discarded by selecting only protons with rigidities $R$ $>$ $R_{min}=10/L^{2}-0.4$ GV, where $L$ is the McIlwain's pa\-ra\-me\-ter \citep{McIlwain}.
Then, each trajectory is back propagated from the measurement location and traced, with no constraint limiting the total path-length or tracing time, until one of the two following conditions is satisfied:
\begin{enumerate}
  \item it reaches the model magnetosphere boundaries (see Section \ref{Geomagnetic field models});
  \item or it reaches an altitude\footnote{Such a value approximately corresponds to the mean production altitude for albedo protons.} of 40 km.
\end{enumerate}
The two categories correspond to ``allowed'' and ``forbidden'' trajectories, respectively:
the former includes contributions from Solar and Galactic CRs (hereafter SCRs and GCRs),
while events satisfying the latter condition,
including albedo particles with rigidities greater than $R_{min}$,
are excluded from the analysis.

\begin{figure}[!t]
\centering
\includegraphics[width=2.8in]{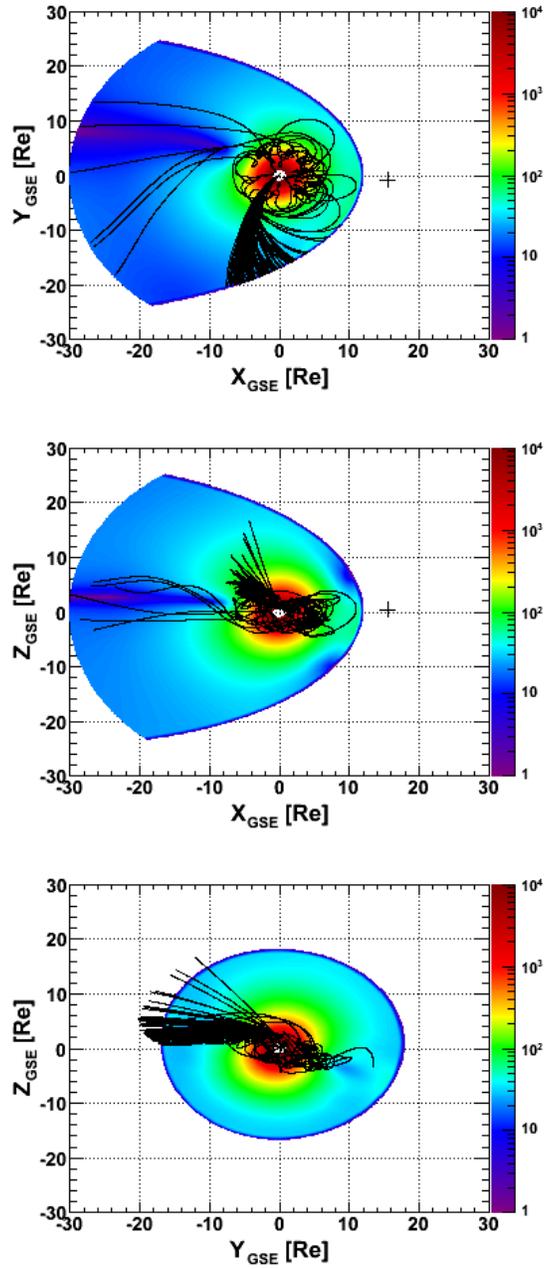}
\caption{Sample CR proton trajectories reconstructed in the magnetosphere. The total magnetic field intensities [nT] are also shown (color coding) for the X-Y (top), the X-Z (middle) and the Y-Z (bottom) GSE planes. See the text for details.}
\label{magnetotraj}
\end{figure}

\section{Asymptotic Arrival Directions}
The asymptotic arrival directions are evaluated with respect to the IMF direction, with polar angles $\alpha$ and $\beta$ denoting the particle pitch and gyro-phase angle, respectively.
Both Geographic (GEO) and Geocentric Solar Ecliptic (GSE) coordinates are used.
Figure \ref{magnetotraj} reports some sample trajectories reconstructed in the GSE coordinate system.
The total magnetic field intensities obtained with the IGRF-11 + TS07D models ($\leq$ 30 $R_{E}$) are also shown (color coding) for the X-Y (top), the X-Z (middle) and the Y-Z (bottom) GSE planes.
The crosses denote the estimated position of the bow shock nose \citep{OMNIWEB}.

\begin{figure*}[!t]
\centering
\includegraphics[width=6.6in]{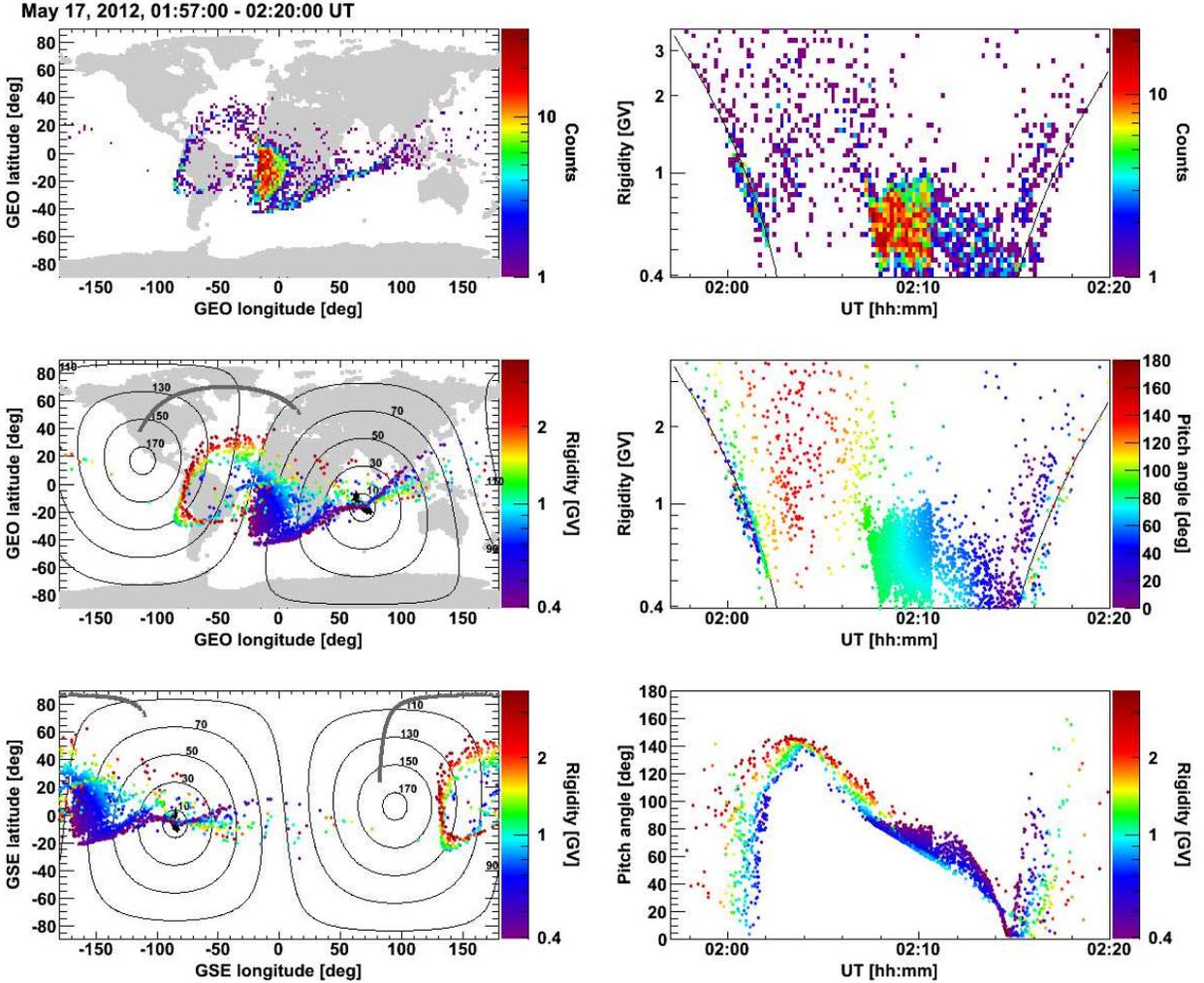}
\caption{Estimated asymptotic arrival directions for protons detected during the May 17, 2012 SEP event. See the text for details.}
\label{17may2012}
\end{figure*}

The trajectory analysis allows a deeper understanding of SEP events.
To improve the interpretation of results, the directions of approach and the entry points at the model magnetosphere boundaries can be visualized as a function of particle rigidity and orbital position \citep{BRUNO-ESTEC}.
As an example, Figure \ref{17may2012} reports the proton results ($R$ $\lesssim$ 3 GV) for the May 17, 2012 SEP event \citep{MAY17PAPER}, associated with the first GLE of the 24$^{rd}$ solar cycle.
Only the first PAMELA polar pass which registered the event is included, corresponding to the interval 01:57 -- 02:20 UT.

Left panels show the reconstructed asymptotic directions for the selected proton sample (counts) in terms of GEO (top and middle) and GSE (bottom) coordinates; in the top panel colors denote the number of proton counts in each bin, while in middle and bottom panels they refer to the particle rigidity. Distributions are integrated over the polar pass. The spacecraft position is indicated by the grey curve.
The contour curves represent values of constant pitch angle with respect to the IMF direction, denoted with crosses.
As PAMELA is moving (eastward) and changing its orientation along the orbit, observed asymptotic directions rapidly vary performing a (clockwise) loop over the region above Brazil (see middle panel).

Right panels in the same figure display the distributions of asymptotic directions as a function of UT, and particle rigidity (top and middle) and pitch-angle (bottom);
colors in the top panel denote the number of proton counts in each bin while, in middle and bottom panel, they correspond to particle pitch-angle and rigidity respectively.
Solid curves denote the estimated St\"{o}rmer vertical cutoff \citep{STORMER,SHEA} for the PAMELA epoch ($\sim 14.3/L^{2}$ GV).

Since PAMELA aperture is about 20 deg, the observable pitch-angle range at a given rigidity is quite small (a few deg) except in the penumbral regions around the local geomagnetic cutoff, where particle trajectories
become complex (chaotic trajectories) and both allowed and forbidden bands of CR trajectories are present \citep{Cooke}.
Corresponding asymptotic directions rapidly change with particle rigidity and looking direction. Conservatively, these regions are excluded from the analysis.

\section{Flux Evaluation}

\subsection{Apparatus Gathering Power}
The factor of proportionality between fluxes and counting rates, corrected for selection efficiencies, is by definition the gathering power $\Gamma$ ($cm^{2}sr$) of the apparatus:
\begin{equation}\label{gathering_power_formula}
\Gamma=\int_{\Omega}d\omega F(\omega) \int_{S} d\textbf{$\sigma\cdot \hat{r}$} = \int_{\Omega} d\omega F(\omega) A(\omega),
\end{equation}
where $\Omega$ is the solid angle domain limited by the instrument geometry, $S$ is the detector surface area, \textbf{$\hat{r}$}d\textbf{$\sigma$} is the effective element of area looking into $\omega$,
$F(\omega)$ is the flux angular distribution (varying between 0 and 1) and $A(\omega)$ is the directional response function \citep{Sullivan}.

\begin{figure}[!t]
\centering
\includegraphics[width=1.8in]{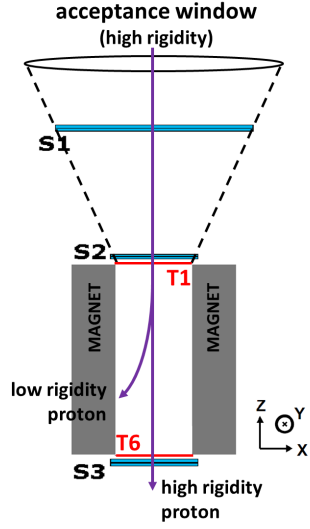}
\caption{Schematic view of the PAMELA apparatus, including only parts constraining the field of view: the Time of Flight system, denoted with S1, S2 and S3; the magnetic spectrometer with first and last tracking system planes (denoted with T1 and T6) and the magnetic cavity. The PAMELA reference frame is also reported.}
\label{Fig1}
\end{figure}

In the case of the PAMELA instrument, $\Gamma$ is rigidity dependent due to the spectrometer bending effect on particle trajectories: it decreases with decreasing rigidity $R$ since particles with lower rigidity are more and more deflected by the magnetic field toward the lateral walls of the magnetic cavity, being absorbed before reaching
the lowest plane of the Time of Flight system, which provides the event trigger (see Figure \ref{Fig1}).

In terms of the zenith $\theta$ and the azimuth $\phi$ angles describing downward-going particle directions in the PAMELA frame\footnote{The PAMELA reference system has the origin in the center of the spectrometer cavity; the Z axis is directed along the main axis of the apparatus, toward the incoming particles; the Y axis is directed opposite to the main direction of the magnetic field inside the spectrometer; the X axis completes a right-handed system.}:
\begin{equation}\label{gathering_power_formula2}
\Gamma(R)=\int_{-1}^{0} dcos \theta \int_{0}^{2\pi} d\phi \hspace{0.04cm} F(R,\theta,\phi)\hspace{0.04cm}   S(R,\theta,\phi) \hspace{0.04cm}  \left|   cos\theta \right|,
\end{equation}
where $S(R,\theta,\phi)$ is the apparatus response function in units of area ($cm^{2}$), and the $\cos\theta$ factor accounts for the trajectory inclination with respect to the instrument axis.

For non-ideal detectors, it is necessary to account for the effect of elastic and inelastic particle interactions inside the apparatus (especially with its mechanical parts),
on the measured coincidence rate.
Consequently, $\Gamma$ is different for each particle species (protons, electrons, ions, etc.).

\subsection{Isotropic Flux Exposition}

For an isotropic particle flux, the gathering power does not depend on looking direction (i.e. $F(\omega)$ = 1), and it is usually called the geometrical factor $G$.

\subsubsection{Monte Carlo Methods}\label{Monte Carlo Integration Methods}
A technically simple but efficient solution for the calculation of the geometrical factor of the apparatus is provided by Monte Carlo methods \citep{Sullivan}.
The solid angle is subdivided into a large number of ($\Delta cos\theta,\Delta\phi$) bins, with the angular domain limited to downward-going directions.
For each bin:
\begin{itemize}
\item particles are produced with random position on a plane generation surface with area $S_{gen}$, placed just above the apparatus opening aperture, so that a large number of events corresponds to the intensity incident on the instrument.
\item For each particle, a random (i.e. random $cos\theta$ and $\phi$) direction is chosen in the selected ($\Delta cos\theta,\Delta\phi$) bin.
\item Trajectories are propagated through the apparatus and tested at successively deeper layers: only particles satisfying all the detector geometrical constraints
    defining the apparatus Field of View (FoV) are selected.
\end{itemize}
The procedure is repeated until the desired statistical precision (see below) is achieved.
Then, for each rigidity value, the geometrical factor is obtained as:
\begin{equation}\label{differential_GF_tot}
G(R) \simeq  S_{gen}\hspace{0.04cm} \Delta cos\theta \hspace{0.04cm} \Delta \phi \sum_{cos\theta} \sum_{\phi} k(R,\theta,\phi)  \hspace{0.04cm} \left| cos\theta\right|
\end{equation}
with
\begin{equation}
k(R,\theta,\phi) =\frac{n_{sel}(R,\theta,\phi)}{n_{tot}(R,\theta,\phi)}, 
\end{equation}%
where $n_{sel}(R,\theta,\phi)$ and $n_{tot}(R,\theta,\phi)$ are the numbers of selected and generated trajectories in each angular bin,
and the $cos\theta$ factor
accounts for the probability of a particle trajectory with direction ($cos\theta,\phi$):
\begin{equation}
P(\theta,\phi)\hspace{0.04cm}dcos\theta\hspace{0.04cm} d\phi = \left|cos\theta\right| \hspace{0.04cm}dcos\theta\hspace{0.04cm} d\phi. 
\end{equation}
The statistical uncertainty on $G(R)$ can be evaluated\footnote{A more rigorous treatment is provided by Bayesian approaches.} with binomial methods:
\begin{equation}\label{stat_differential_error}
\Delta G(R) \simeq S_{gen} \hspace{0.04cm}\Delta cos\theta \hspace{0.04cm} \Delta \phi   \sqrt{ \sum_{cos\theta} \sum_{\phi} \left[ \frac{k(R,\theta,\phi)(1-k(R,\theta,\phi))}{n_{tot}(R,\theta,\phi)}\right]  cos^{2}\theta}.
\end{equation}
Advantage is taken of the angular subdivision, by varying directions over the $\Delta cos\theta\Delta\phi$ solid angle rather than a
full $2\pi$ hemisphere. In particular, a number $n_{tot}(R,\theta,\phi)$ of incident trajectories proportional to $cos^{-1}\theta_{c}$ (with $\theta_{c}$ taken at $\Delta cos\theta$ bin center) is chosen
in order to minimize the statistical error associated to the angular bins with a small $k(R,\theta,\phi)$.

\begin{figure}[!t]
\centering
\includegraphics[width=4.5in]{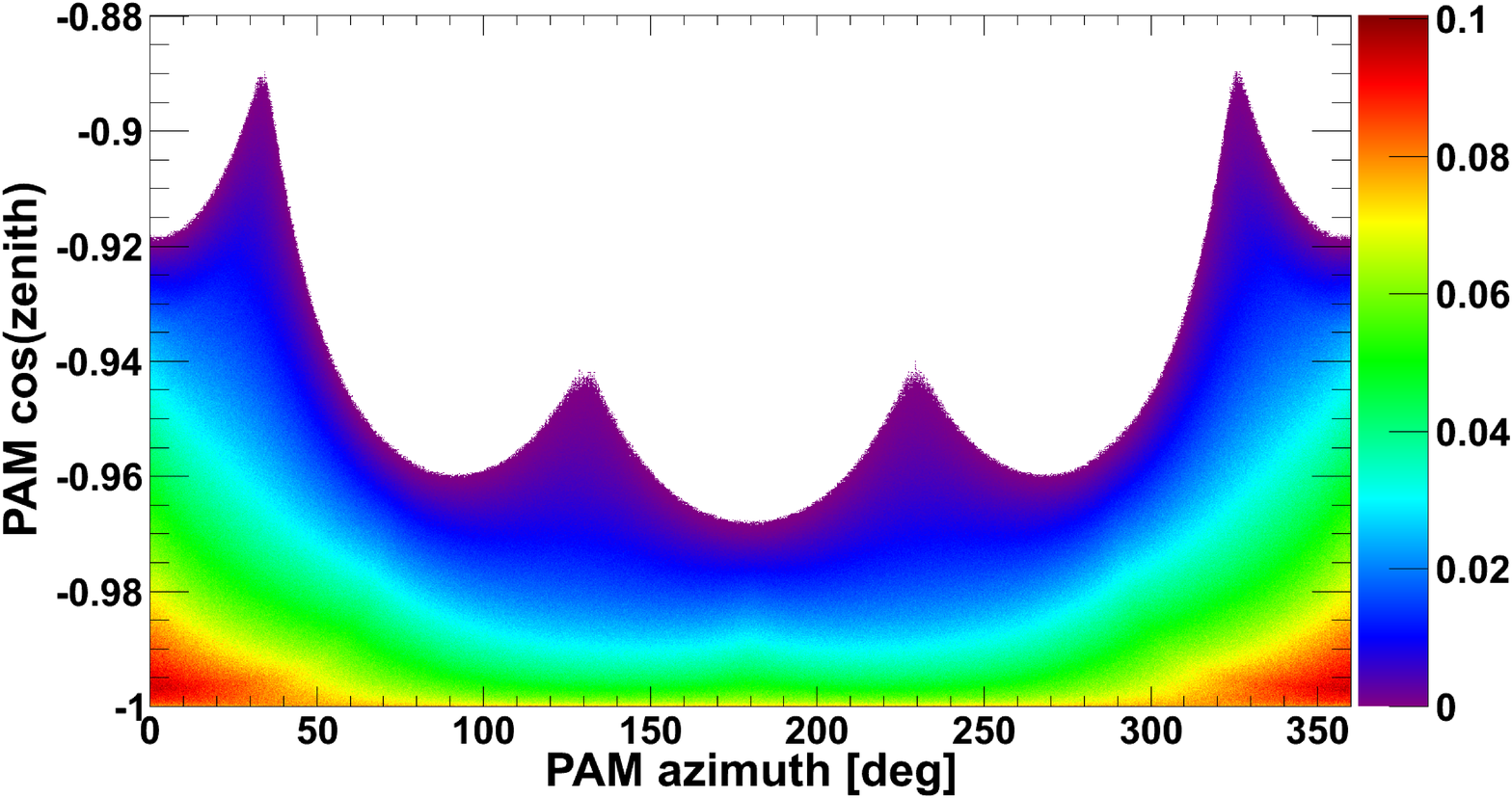}\\
\includegraphics[width=4.5in]{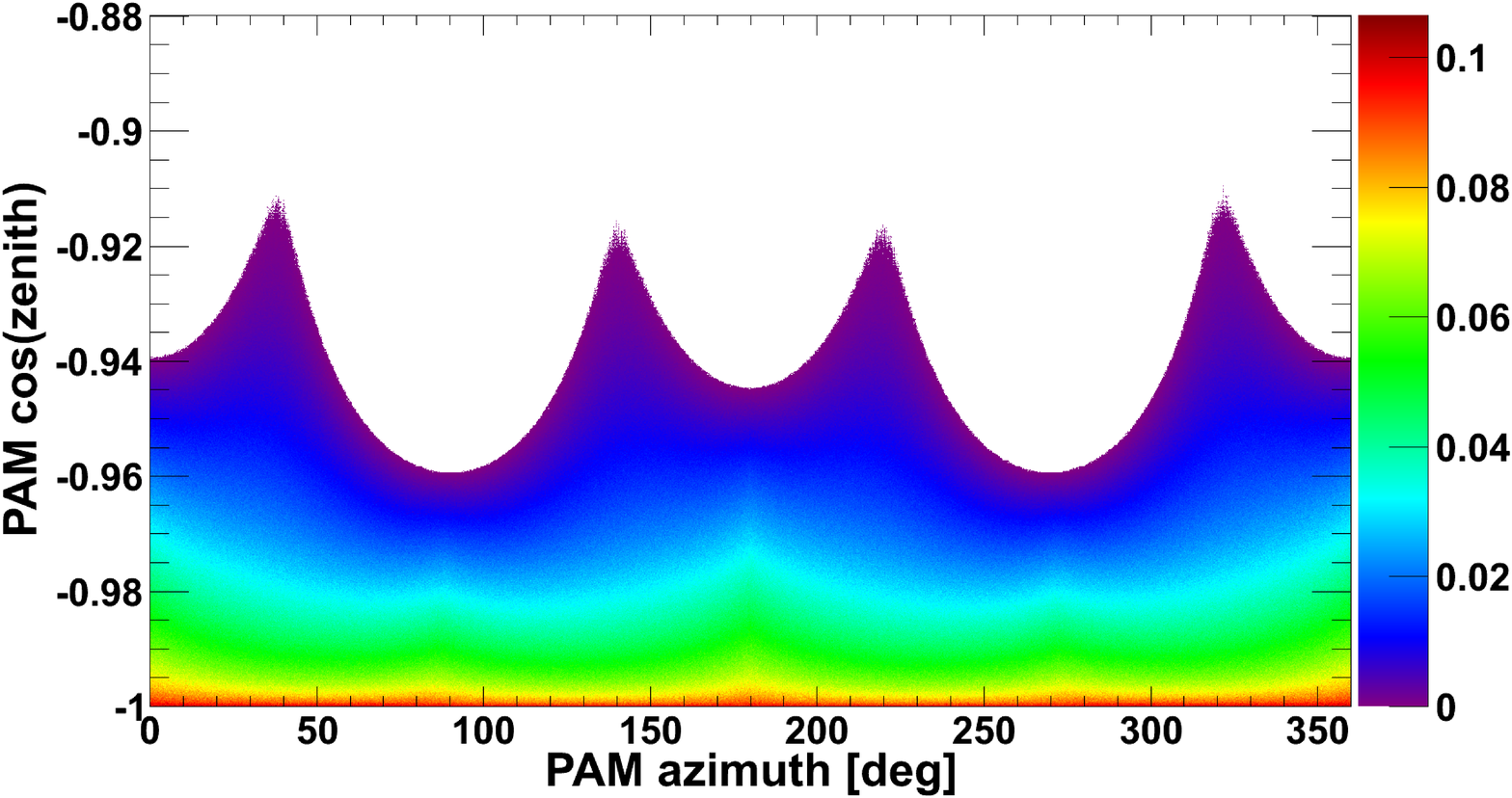}
\caption{The $k(R,\theta,\phi)= n_{sel}(R,\theta,\phi) / n_{tot}(R,\theta,\phi)$ ratio (color codes) over the PAMELA field of view, as a function of local polar coordinates $\phi$ and $cos\theta$ (apparatus reference frame). Results for 0.39 GV (top) and 4.09 GV (bottom) protons are displayed. See the text for details.}
\label{acce78}
\end{figure}

The dependence of the instrument response on particle rigidity is studied by performing the calculation for 30 rigidities in the range 0.39--10 GV, and the value of $G(R)$ at intermediate $R$ is evaluated through interpolation methods.
An accurate estimate of the PAMELA geometrical factor based on the Monte Carlo approach can be found in \citet{BRUNO_PHD}.

As an example, Figure \ref{acce78} reports the $k(R,\theta,\phi)$ ratio (color codes) over the PAMELA FoV as a function of local $\phi$ and $cos\theta$, for 0.39 GV (top panel) and 4.09 GV (bottom panel) protons.
The four peaks structure reflects the rectangular section of the apparatus; the differences in the FoV distributions at different rigidities are due to the bending effect of the magnetic spectrometer.

\subsection{Anisotropic Flux Exposition}
However, in presence of an anisotropic flux exposition ($F\ne const$), the gathering power depends also on the flux angular distribution.
SCR fluxes can be conveniently expressed in terms of asymptotic polar angles $\alpha$ (pitch angle) and $\beta$ (gyro-phase angle) with respect to the IMF direction: $F=F(R,\alpha,\beta)$. The corresponding gathering power can be written as:
\begin{equation}\label{gathering_power_formula4}
\Gamma(R)=\int_{0}^{\pi} sin\alpha \hspace{0.04cm} d\alpha  \int_{0}^{2\pi} d\beta  \hspace{0.04cm}   F(R,\alpha,\beta) \hspace{0.04cm} S(R,\theta,\phi) \hspace{0.04cm}\left|  cos\theta \right|,
\end{equation}
with $\theta$=$\theta(R,\alpha,\beta)$ and $\phi$=$\phi(R,\alpha,\beta)$. The flux angular distribution $F(R,\alpha,\beta)$ is unknown.

\subsubsection{Asymptotic Effective Area}
For simplicity, we assume that SCR fluxes depend only on particle rigidity $R$ and asymptotic pitch angle $\alpha$, estimating an apparatus effective area (cm$^{2}$) as:
\begin{equation}\label{gathering_power_formula5}
H(R,\alpha)=\frac{sin\alpha}{2\pi}\int_{0}^{2\pi} d\beta  \hspace{0.04cm} S(R,\theta,\phi)\hspace{0.04cm}  \left|  cos\theta \right|,
\end{equation}
by averaging the directional response function over the $\beta$ angle.
The method is valid also for isotropic fluxes (independent on $\alpha$): in this case, the effective area is related to the geometrical factor $G(R)$ by:
\begin{equation}\label{gathering_power_formula6}
G(R)=2\pi  \int_{0}^{\pi}d\alpha \hspace{0.04cm}H(R,\alpha).
\end{equation}

The approach is analogous to the one developed for the measurement of geomagnetically trapped protons \citep{BRUNO_ICRC_SUBCUTOFF}, with $\alpha$ and $\beta$ denoting the polar angles with respect to the local geomagnetic field direction. But while local angles can be calculated from $(\theta, \phi)$ by means of basic rotation matrices of the FoV involving only the spacecraft orientation with respect to the local magnetic field,
the conversion to asymptotic angles depends on the particle propagation in the magnetosphere, so trajectory tracing methods are needed.

\subsubsection{Area Calculation}\label{Area Calculation}

The effective area definition given in Equation \ref{gathering_power_formula5} is based on the assumption of approximately isotropic fluxes within small pitch-angle bins.
Consequently, $H(R,\alpha)$ can be derived from Equation \ref{differential_GF_tot} by integrating
the directional response function
over the ($cos\theta,\phi$) directions
corresponding
to pitch angles within the interval $\alpha\pm\Delta\alpha/2$:
\begin{equation}\label{MCgfactor_alpha}
2\pi  \int_{\Delta\alpha}d\alpha \hspace{0.04cm}H(R,\alpha) \simeq S_{gen}  \hspace{0.04cm}\Delta cos\theta \hspace{0.04cm} \Delta \phi   \sum_{\theta,\phi\rightarrow\alpha}  k(R,\theta,\phi) \hspace{0.04cm}\left|cos\theta \right|.
\end{equation}
The approach accuracy depends on the number of ($\Delta cos\theta,\Delta\phi$) bins used in the angular partitioning (see Section \ref{Monte Carlo Integration Methods}), while the width of
the $\Delta\alpha$ bins is chosen accounting for the detector angular resolution.

\begin{figure}[!t]
\centering
\includegraphics[width=5in]{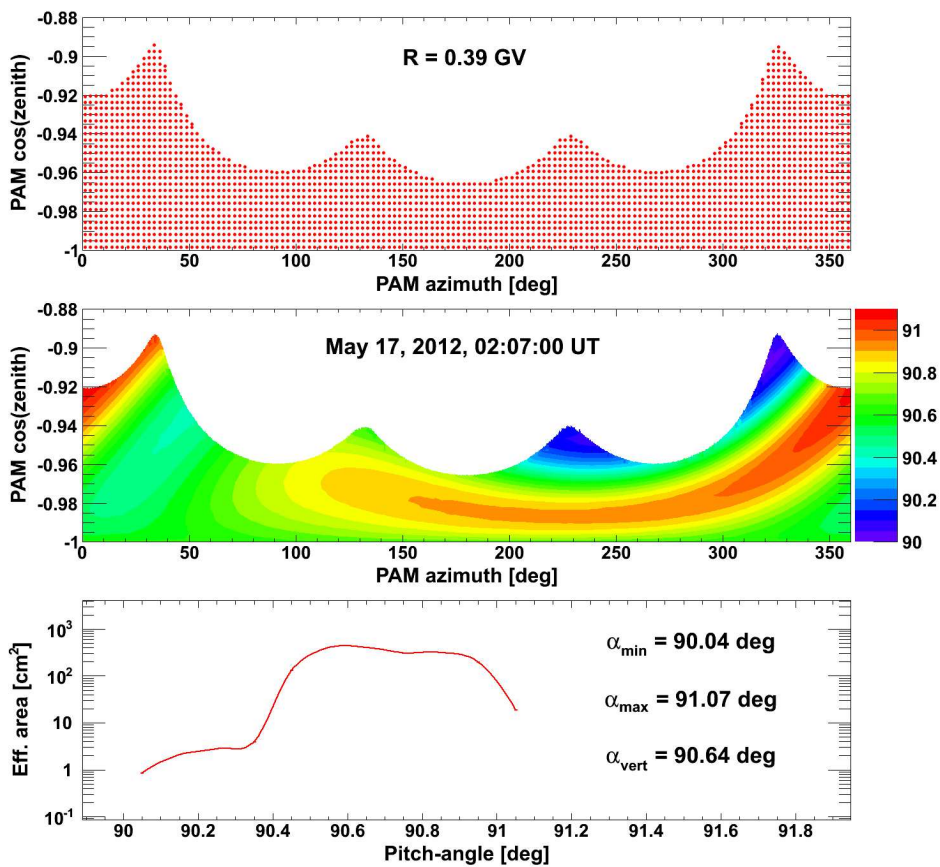}
\caption{Top: simulated directions (red points) inside PAMELA field of view. Middle: pitch-angle coverage (color axis, deg). Bottom: the apparatus effective area as function of pitch-angle; minimum and maximum observable pitch-angles are reported, along with the value corresponding to the vertical direction. Results correspond to 0.39 GV protons for a sample orbital position (May 17, 2012, 02:07:00 UT).}
\label{area600_ek78}
\end{figure}

\begin{figure}[!t]
\centering
\includegraphics[width=5in]{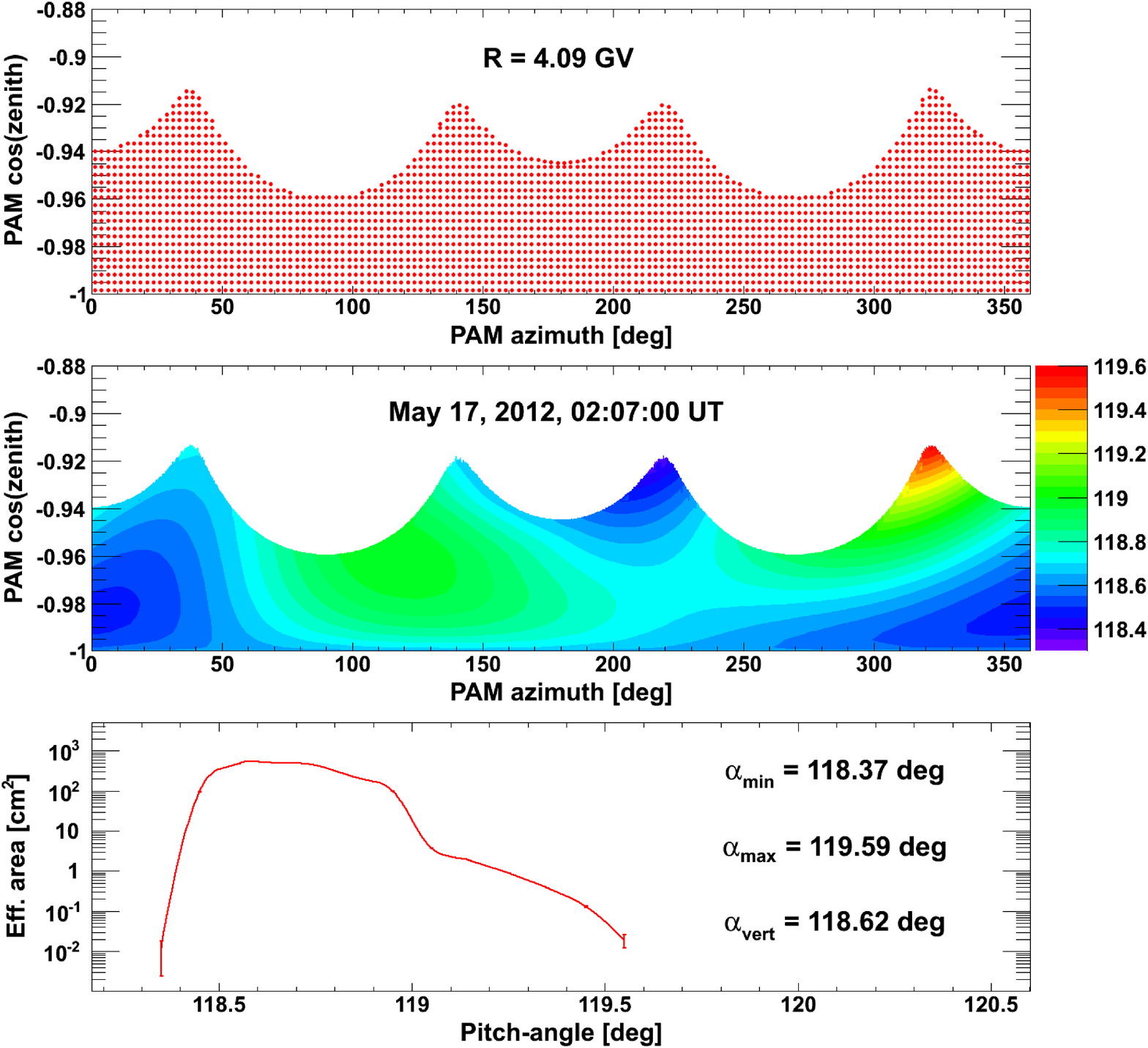}
\caption{Same than Figure \ref{area600_ek78} but for 4.09 GV protons.}
\label{area600_ek1582}
\end{figure}

To convert local ($cos\theta,\phi$) into asymptotic ($\alpha,\beta$) directions and apply Equation \ref{MCgfactor_alpha}, a large number of trajectories $N$, uniformly distributed inside PAMELA field of view, has to be reconstructed in the magnetosphere, for each rigidity value and each orbital position.
In order to assure a high resolution,
the calculation is performed for time steps with a 1-sec width,
back-tracing
about 2800 trajectories for each rigidity bin,
for a total of more than $10^{8}$ trajectories for each polar pass ($\sim$ 23 min).
At a later stage, results are extended over the full
field of view of PAMELA
through a bilinear interpolation.

\begin{figure}[!t]
\centering
\includegraphics[width=5.5in]{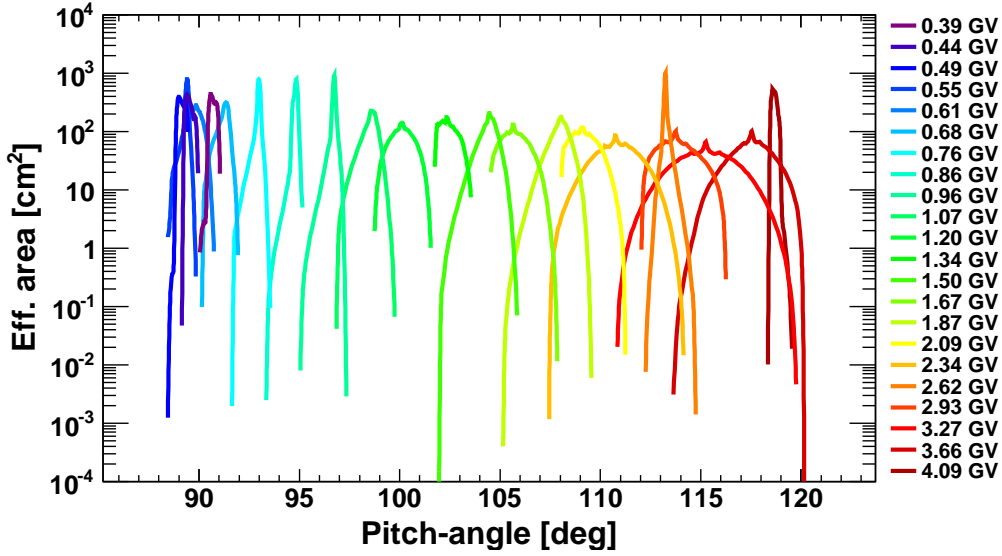}
\caption{The PAMELA effective area at a sample orbital position (May 17, 2012, 02:07:00 UT) as function of pitch-angle, for different values of particle rigidity.}
\label{all600}
\end{figure}

The procedure is illustrated in Figures \ref{area600_ek78}-\ref{area600_ek1582} for 0.39 and 4.09 GV protons respectively, at a sample orbital position (May 17, 2012, 02:07 UT).
Top panels report the distribution of reconstructed directions in the PAMELA field of view, with each point as\-so\-cia\-ted to a given asymptotic direction ($\alpha$,$\beta$); middle panels show the calculated (after interpolation) pitch-angle coverage\footnote{The $\beta$ calculation can be neglected under the gyro-tropic approximation.};
bottom panels illustrate the estimated effective area as a function of
the explored pitch-angle range.
Final calculation results for 22 ri\-gi\-di\-ty values between 0.39 and 4.09 GV (see the color code) are displayed in Figure \ref{all600}: the peaks in the distributions
correspond to vertically incident protons.

\begin{figure}[!t]
\centering
\includegraphics[width=5.5in]{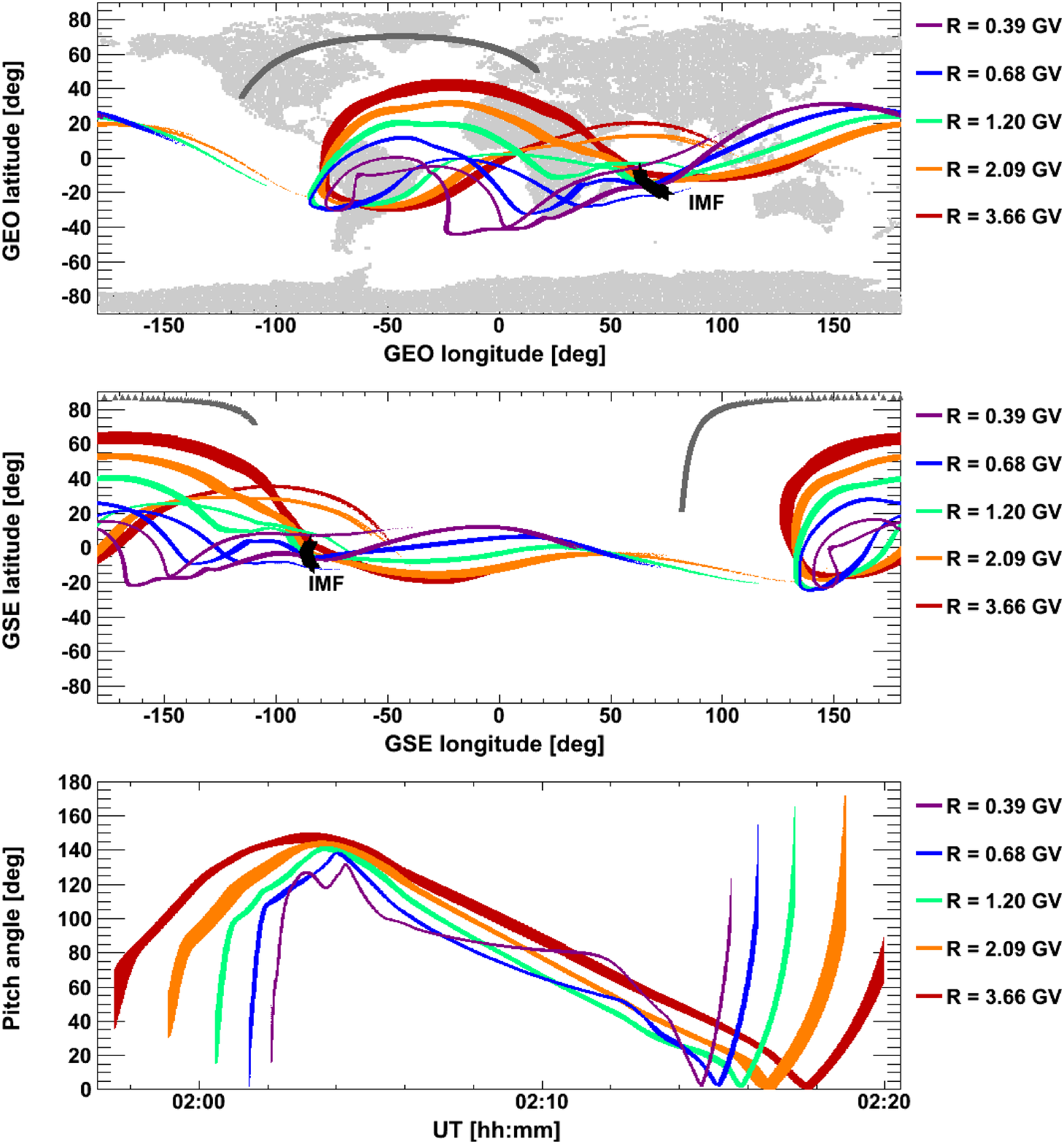}
\caption{Asymptotic cones of acceptance of the PAMELA apparatus for sample rigidity values (see labels), evaluated in GEO (top) and GSE (middle) coordinates, and as a function of UT and pitch-angle (bottom). Grey points denote the spacecraft position, while crosses indicate the IMF direction. Calculations refer to the first PAMELA polar pass (01:57--02:20 UT) during the May 17, 2012 SEP event.}
\label{acce_cones2012}
\end{figure}

Figure \ref{acce_cones2012} reports the asymptotic cones of acceptance of the PAMELA apparatus
evaluated for the first polar pass (01:57--02:20 UT) during the May 17, 2012 SEP event \citep{MAY17PAPER}.
Results for sample rigidity values (see the color code)
are shown as a function of GEO (top panel) and GSE (middle panel) coordinates;
grey points denote the spacecraft position, while crosses indicate the IMF direction.
The pitch-angle coverage as a function of orbital position is shown in the bottom panel.
While moving (and rotating) along the orbit, PAMELA covers a large pitch-angle interval, approximately ranging from 0 to 145 deg.
In particular, PAMELA is looking at the IMF direction between 02:14 and 02:18 UT, depending on the proton rigidity.

\subsection{Differential Directional Fluxes}
Differential directional fluxes are obtained
at each orbital position $t$ as:
\begin{equation}\label{eq_flux}
\Phi(R,\alpha,t) = \frac{N_{tot}(R,\alpha,t)}{ 2\pi\int\limits_{\Delta R} dR \int\limits_{\Delta \alpha} d\alpha \int\limits_{\Delta t}^{} dt H(R,\alpha,t) },
\end{equation}
where
$N_{tot}(R,\alpha,t)$ is the number of proton counts in the bin $(R,\alpha,t)$, corrected by the selection efficiencies, and
the denominator represents the asymptotic exposition of the apparatus integrated over the selected rigidity bin $\Delta R$.

Averaged fluxes over the polar pass $T=\sum\Delta t$ are evaluated as:
\begin{equation}\label{eq_flux_mean}
\Phi(R,\alpha) =  \frac{N_{tot}(R,\alpha)}{ 2\pi\int\limits_{\Delta R} dR \int\limits_{\Delta \alpha} d\alpha \int\limits_{T}^{} dt H(R,\alpha,t) },
\end{equation}
where $N_{tot}(R,\alpha)=\sum_{T}^{} N_{tot}(R,\alpha,t)$ and the exposition is derived by weighting each effective area contribution by the corresponding lifetime spent by PAMELA at the same orbital position.

\subsection{GCR Background Subtraction}
Since it is not possible to discriminate between SCR and GCR signals, solar flux intensities are obtained by subtracting the GCR contribution from the total measured flux.
The GCR component is evaluated by estimating proton fluxes during two days prior the arrival of SEPs. We found that the GCR flux is isotropic with respect to the IMF direction within experimental uncertainties. Consequently, the same flux $\Phi_{GCR}(R)$ is subtracted for all pitch angle bins:
\begin{equation}\label{eq_flux_sub}
\begin{split}
\Phi_{SCR}(R,\alpha) =  \Phi_{tot}(R,\alpha) - \Phi_{GCR}(R)  =  \\
= \frac{N_{tot}(R,\alpha)-N_{GCR}(R,\alpha)}{ 2\pi\int\limits_{\Delta R} dR \int\limits_{\Delta \alpha} d\alpha \int\limits_{T}^{} dt H(R,\alpha,t) }.
\end{split}
\end{equation}

\subsection{Flux Uncertainties}
Flux statistical errors can be obtained by evaluating 68.27\% C.L. intervals for a poissonian signal $N_{tot}(R,\alpha)$ in presence of a background $N_{GCR}(R,\alpha)$ \citep{FELDMAN}.
Systematic uncertainties related to the reconstruction of asymptotic directions are evaluated by introducing a bias in the particle direction measurement from the tracking system, according to a gaussian distribution with a variance equal to the experimental angular resolution.

\section{Conclusions}\label{Conclusions}
This paper reports the analysis methods developed for the estimate of SEP energy spectra as a function of the particle asymptotic direction of arrival.
The exposition of the PAMELA apparatus is evaluated by means of accurate back-tracing simulations based on a realistic description of the Earth's magnetosphere.
As case study, the results of the calculation for the May 17, 2012 event are discussed. The developed trajectory analysis enables the investigation of flux anisotropies, pro\-viding fundamental information for the characterization of SEPs. It will prove to be a vital in\-gre\-dient for the interpretation of solar events observed by PAMELA during solar cycles 23 and 24.

\section*{Acknowledgements}
We acknowledge support from The Italian Space Agency (ASI), Deutsches Zentrum f$\ddot{u}$r Luftund Raumfahrt (DLR), The Swedish National Space Board, The Swedish Research Council, The Russian Space Agency (Roscosmos) and The Russian Scientific Foundation.
We gratefully thank N. Tsyganenko for helpful discussions, and M. I. Sitnov and G. K. Stephens for their assistance and support in the use of the TS07D model.

\end{document}